# Piezooptic Properties of α-BaB$_2$O$_4$ and Li$_2$B$_4$O$_7$ Crystals

Martunyuk-Lototska I., Mys O., Dyachok Ya., Dudok T., Adamiv V., Burak Ya. and Vlokh R.

Institute of Physical Optics, 23 Dragomanov Str., 79005 Lviv, Ukraine

## Abstract

Photoelastic coefficients of α-BaB$_2$O$_4$ and Li$_2$B$_4$O$_7$ crystals are calculated on the basis of piezooptic measurements performed with interferometric technique and the elastic compliance and stiffness data. Using the experimental results, the acoustooptic (AO) figure of merit (FM) has been estimated for the possible geometries of AO interaction. It is shown that the AO FM for the ABO and LTB crystals reach respectively the values $M_2$=243.4×10$^{-15}$s$^3$/kg and $M_2$=2.57×10$^{-15}$s$^3$/kg, if the interaction with the slowest ultrasonic waves ($v$=933.5m/s and $v$=3173m/s) is concerned. The directions of propagation and polarization of those acoustic waves are obtained on the basis of construction of indicative surfaces of the acoustic wave velocities.

**Key words:** piezooptic effect, borate crystals, photoelastic effect

**PACS** 78.20.Hp

## Introduction

α-BaB$_2$O$_4$ and Li$_2$B$_4$O$_7$ crystals belong to borate crystal family and are described respectively with the point symmetry groups $\bar{3}$m and 4mm. Acentric borate crystals are widely used as nonlinear optical materials due to high values of their nonlinear susceptibilities [1,2] and a high level of optical damage threshold [3]. In our previous papers [4,5] we have reported that, for example, β-BaB$_2$O$_4$ is characterized with quite low transverse acoustic wave velocities and could be therefore used as a promising acoustooptic (AO) material. The value of AO figure of merit (abbreviated hereafter as AOFM) for β-BaB$_2$O$_4$ crystals [5] calculated on the basis of photoelastic coefficients, refractive indices and the ultrasonic wave velocities is comparable with those typical for the well-known AO materials such as lithium niobate or Pb$_2$MoO$_5$ [6]. However, the growing process for β-BaB$_2$O$_4$ crystals is time-consuming, when compare to α-BaB$_2$O$_4$ and Li$_2$B$_4$O$_7$. On the other side, the photoelastic parameters that affect the AOFM of α-BaB$_2$O$_4$ and Li$_2$B$_4$O$_7$ crystals are still unknown (to our knowledge, only $p_{66}$ coefficient has been determined for Li$_2$B$_4$O$_7$ crystals [7]). This is why we report below the results of studies of piezooptic effect in these crystals, together with the estimation of AOFM.

## Experimental results

α-BaB$_2$O$_4$ and Li$_2$B$_4$O$_7$ crystals were grown with the Czochralski method. Single crystals of a good optical quality with 3×3×3 cm$^3$ dimensions were obtained after a one-week growing process. Their piezooptic coefficients were measured at room temperature with interferometric technique using the Mach-Zender interferometer ($\lambda$=632.8nm). For avoiding ambiguity in the presentation of results, the elastic contribution was derived with the aid of the relationship $\delta(\Delta nd)_{ij} = \pi_{ijkl}\sigma_{kl} - (n_c - 1)S_{ijkl}\sigma_{kl}$. The ultrasonic wave velocities were measured at room temperature with the pulse-echo overlap method [8]. The accuracy for the absolute velocity values was about 0.5%. The acoustic waves in samples were excited with LiNbO$_3$ transducers characterized with the resonance frequency of $f$ = 10 MHz, the bandwidth of $\Delta f$ = 0.1 MHz and the acoustic power from $P_a$=1 to 2W. The photoelastic coefficients were calculated on the basis of elastic stiffness data obtained earlier [9], using the known formula $p_{\lambda\mu} = \pi_{\lambda\nu}C_{\nu\mu}$.

The results for the piezooptic coefficients of Li$_2$B$_4$O$_7$ crystals are presented below in the form of matrix:

$$\pi_{\lambda\nu} = \begin{pmatrix} \pi_{11} & \pi_{12} & \pi_{13} & 0 & 0 & 0 \\ \pi_{12} & \pi_{11} & \pi_{13} & 0 & 0 & 0 \\ \pi_{31} & \pi_{31} & \pi_{33} & 0 & 0 & 0 \\ 0 & 0 & 0 & \pi_{44} & 0 & 0 \\ 0 & 0 & 0 & 0 & \pi_{44} & 0 \\ 0 & 0 & 0 & 0 & 0 & \pi_{66} \end{pmatrix} = \begin{pmatrix} -2.75 & -0.45 & 1.08 & 0 & 0 & 0 \\ -0.45 & -3.52 & 1.08 & 0 & 0 & 0 \\ -2.53 & -2.53 & 2.9 & 0 & 0 & 0 \\ 0 & 0 & 0 & 1.1 & 0 & 0 \\ 0 & 0 & 0 & 0 & 1.1 & 0 \\ 0 & 0 & 0 & 0 & 0 & 1.05 \end{pmatrix} \times 10^{-12} m^2/N$$

while for α-BaB$_2$O$_4$ we have

$$\pi_{\lambda\nu} = \begin{pmatrix} \pi_{11} & \pi_{12} & \pi_{13} & \pi_{14} & 0 & 0 \\ \pi_{12} & \pi_{11} & \pi_{13} & -\pi_{14} & 0 & 0 \\ \pi_{31} & \pi_{31} & \pi_{33} & 0 & 0 & 0 \\ \pi_{41} & \pi_{41} & 0 & \pi_{44} & 0 & 0 \\ 0 & 0 & 0 & 0 & \pi_{44} & 2\pi_{41} \\ 0 & 0 & 0 & 0 & \pi_{14} & \pi_{66} \end{pmatrix} = \begin{pmatrix} 1.58 & 2.08 & -4.32 & -14.22 & 0 & 0 \\ 2.08 & 1.58 & -4.32 & 14.22 & 0 & 0 \\ 1.52 & 1.52 & -7.24 & 0 & 0 & 0 \\ 2.85 & 2.85 & 0 & -24.58 & 0 & 0 \\ 0 & 0 & 0 & 0 & -24.58 & 5.7 \\ 0 & 0 & 0 & 0 & -14.22 & 0.5 \end{pmatrix} \times 10^{-12} m^2/N$$

The calculated values of photoelastic coefficients for Li$_2$B$_4$O$_7$ crystals are as follows:

$$p_{\lambda\mu} = \begin{pmatrix} p_{11} & p_{12} & p_{13} & 0 & 0 & 0 \\ p_{12} & p_{11} & p_{13} & 0 & 0 & 0 \\ p_{31} & p_{31} & p_{33} & 0 & 0 & 0 \\ 0 & 0 & 0 & p_{44} & 0 & 0 \\ 0 & 0 & 0 & 0 & p_{44} & 0 \\ 0 & 0 & 0 & 0 & 0 & p_{66} \end{pmatrix} = \begin{pmatrix} -0.32 & -0.04 & -0.06 & 0 & 0 & 0 \\ -0.04 & -0.32 & -0.06 & 0 & 0 & 0 \\ -0.24 & -0.24 & -0.02 & 0 & 0 & 0 \\ 0 & 0 & 0 & 0.05 & 0 & 0 \\ 0 & 0 & 0 & 0 & 0.05 & 0 \\ 0 & 0 & 0 & 0 & 0 & 0.06 \end{pmatrix}$$

The same values for α-BaB$_2$O$_4$ crystals may be written as

$$p_{\lambda\mu} = \begin{pmatrix} p_{11} & p_{12} & p_{13} & p_{14} & 0 & 0 \\ p_{12} & p_{11} & p_{13} & -p_{14} & 0 & 0 \\ p_{31} & p_{31} & p_{33} & 0 & 0 & 0 \\ p_{41} & -p_{41} & 0 & p_{44} & 0 & 0 \\ 0 & 0 & 0 & 0 & p_{44} & p_{41} \\ 0 & 0 & 0 & 0 & p_{14} & p_{66} \end{pmatrix} = \begin{pmatrix} 0.03 & 0.16 & 0.14 & 0.06 & 0 & 0 \\ 0.16 & 0.03 & 0.14 & -0.06 & 0 & 0 \\ -0.15 & -0.15 & -0.16 & 0 & 0 & 0 \\ -0.05 & 0.05 & 0 & -0.1 & 0 & 0 \\ 0 & 0 & 0 & 0 & -0.1 & -0.05 \\ 0 & 0 & 0 & 0 & 0.06 & -0.07 \end{pmatrix}$$

## Discussion

Basing upon the photoelastic coefficients determined above, together with the ultrasonic velocities, crystal density and the refractive indices, one can estimate the AOFM for different geometries of AO interaction, under the circumstances that the incident light and the acoustic wave propagate along the principal axes of optical indicatrix (see Table 1a,b).

It is interesting to note that the Bragg conditions can be satisfied only in a few cases of anisotropic AO interaction. The maximum value of AOFM ($M_2$=32.971×10$^{-15}$s$^3$/kg) refers to the ABO crystals and corresponds to the case of interaction with the slowest acoustic wave having the direction of propagation [100] and that of polarization [001] (see Figure 1).

On the other hand, it follows from the calculated indicative surfaces of ultrasonic velocities that the minimum value of $v$=933.5m/s for the ABO crystals corresponds to the acoustic wave with the *k*-vector lying in (011) plane and making the angle 18° with respect to *z* axis and the projections of the unit displacement vector $X_z$=0.457, $X_y$=-0.899 and $X_x$=0 (see Figure 2a). The LTB crystals exhibit the two such directions of propagation for the slowest waves, with the same velocities ($v$=3173m/s), i.e. we have the cases:

Table 1a. AO parameters of the LTB crystals ($\rho$=2420kg/m$^3$, $n_o$=1.6084 and $n_e$=1.5516).

| Acoustic wave | | p | \|p$_{eff}$\| | n | Light | | $M_2$,10$^{-15}$, s$^3$/kg or possibility for matching the Bragg conditions |
|---|---|---|---|---|---|---|---|
| V, m/s | Propagation direction, Polarization | | | | Direction | Polarization | |
| 7358 | [100], [100] | p$_{11}$ | 0.32 | n$_e$ | [010] | [100] | not |
| | | p$_{31}$ | 0.24 | n$_o$ | | [001] | not |
| | | p$_{11}$ | 0.32 | n$_o$ | [001] | [100] | not |
| | | p$_{12}$ | 0.04 | n$_o$ | | [010] | not |
| 7460 | [010], [010] | p$_{11}$ | 0.32 | n$_e$ | [100] | [010] | not |
| | | p$_{31}$ | 0.24 | n$_o$ | | [001] | not |
| | | p$_{12}$ | 0.04 | n$_o$ | [001] | [100] | not |
| | | p$_{11}$ | 0.32 | n$_o$ | | [010] | not |
| 5036 | [001], [001] | p$_{13}$ | 0.06 | n$_e$ | [100] | [010] | not |
| | | p$_{33}$ | 0.02 | n$_o$ | | [001] | not |
| | | p$_{13}$ | 0.06 | n$_e$ | [010] | [100] | not |
| | | p$_{33}$ | 0.02 | n$_o$ | | [001] | not |
| 4448 | [100], [010] | p$_{66}$ | 0.06 | n$_o$ | [001] | [100], [010] | not |
| 4769 | [100], [001] | p$_{44}$ | 0.05 | n$_e$ | [010] | [100], [001] | 0.134 |
| 4610 | [010], [001] | p$_{44}$ | 0.05 | n$_e$ | [100] | [010], [001] | 0.149 |

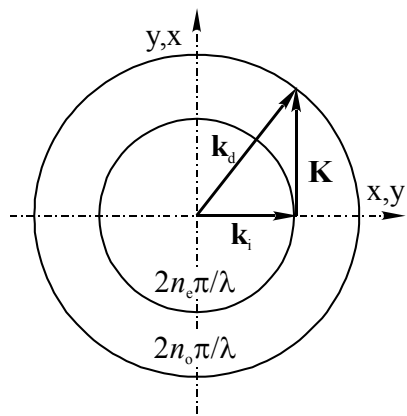

**Fig. 1.** Diagram of AO interaction in the ABO crystals for the case of propagation of acoustic wave and the incident optical wave along one of the principal axis of optical indicatrix (in this particular case the value $M_2$=32.971×10$^{-15}$s$^3$/kg can be achieved).

1) *k*-vector lies in (011) plane at the angle of 39° with respect to *z* axis; the projections of the unit displacement vector are $X_z$=0.794, $X_y$=-0.606 and $X_x$=0 (Figure 2b);
2) *k*-vector lies in (101) plane at the angle of 54° with respect to *z* axis; the projections of the unit displacement vector are $X_z$=0.79, $X_y$=0 and $X_x$=-0.61 (Figure 2b).

Let us evaluate the AOFM for the both cases mentioned above. For this aim one should derive the expression for the effective photoelastic coefficient $p_{eff}$. Let us, for instance, consider the ABO crystals and the incident optical wave propagated along the [010] direction with $E_3$ polarization. The acoustic wave (according to the Bragg condition, the acoustic frequency should be equal to $f_a$=3×10$^9$Hz) is propagated in the (011) plane at the angle of 18° with respect to z axis and the projections of the unit displacement vector are as follows: $X_z$=0.457, $X_y$=-0.899 and $X_x$=0. Then the optical indicatrix equation may be written as

$$(B_1 + p_{12}e_2 + p_{13}e_3 + p_{14}e_4)x^2 +$$
$$+(B_1 + p_{11}e_2 + p_{13}e_3 - p_{14}e_4)y^2 +$$
$$(B_3 + p_{31}e_2 + p_{33}e_3)z^2 +$$
$$+2(p_{44}e_4 - p_{41}e_2)yz = 1 \qquad (1)$$

where $B_i$ are the optical impermeability constants and $e_j$ the strains induced by the acoustic wave. After rewriting Eq. (1) in the proper coordinate system of crystal, we obtain

$$(B_1 + p_{12}e_2 + p_{13}e_3 + p_{14}e_4)X^2 +$$
$$(B_1 + p_{11}e_2 + p_{13}e_3 - p_{14}e_4 + \frac{(p_{44}e_4 - p_{41}e_2)^2}{(B_1 - B_3 + (p_{11} - p_{31})e_2 + (p_{13} - p_{33})e_3 - p_{14}e_4)})Y^2 + \quad (2)$$
$$(B_3 + p_{31}e_2 + p_{13}e_3 - \frac{(p_{44}e_4 - p_{41}e_2)^2}{(B_1 - B_3 + (p_{11} - p_{31})e_2 + (p_{13} - p_{33})e_3 - p_{14}e_4)})Z^2 = 1$$

Table 1b. AO parameters of the ABO crystals ($\rho$=3747kg/m$^3$, $n_o$=1.667 and $n_e$=1.528).

| Acoustic wave | | p | \|p$_{eff}$\| | n | Light | | M$_2$, 10$^{-15}$ s$^3$/kg or possibility for matching the Bragg conditions |
|---|---|---|---|---|---|---|---|
| V, m/s | Propagation direction, Polarization | | | | Direction | Polarization | |
| 5649 | [100], [100] | p$_{11}$ | 0.03 | n$_e$ | [010] | [100] | not |
| | | p$_{31}$ | 0.15 | n$_o$ | | [001] | not |
| | | p$_{21}$ | 0.16 | n$_o$ | [001] | [010] | not |
| | | p$_{11}$ | 0.03 | n$_o$ | | [100] | not |
| 5437 | [010], [010] | p$_{11}$ | 0.03 | n$_e$ | [100] | [010] | not |
| | | p$_{42}$=-p$_{41}$ | 0.05 | n$_o$ | | [001], [010] | 0.075 |
| | | p$_{31}$ | 0.15 | n$_o$ | | [001] | not |
| | | p$_{42}$=-p$_{41}$ | 0.05 | n$_e$ | | [010], [001] | 0.045 |
| | | p$_{12}$ | 0.16 | n$_o$ | [001] | [100] | not |
| | | p$_{22}$=p$_{11}$ | 0.03 | n$_o$ | | [010] | not |
| 3221 | [001], [001] | p$_{23}$=p$_{13}$ | 0.14 | n$_e$ | [100] | [010] | not |
| | | p$_{33}$ | 0.16 | n$_o$ | | [001] | not |
| | | p$_{13}$ | 0.14 | n$_e$ | [010] | [100] | not |
| | | p$_{33}$ | 0.16 | n$_o$ | | [001] | not |
| 2959 | [100], [010] | p$_{56}$=p$_{41}$ | 0.05 | n$_o$ | [010] | [001], [100] | 0.468 |
| | | p$_{41}$=p$_{56}$ | 0.05 | n$_e$ | | [100], [001] | 0.277 |
| | | p$_{66}$ | 0.07 | n$_o$ | [001] | [010], [100] | not |
| | | p$_{66}$ | 0.07 | n$_o$ | | [100], [010] | not |
| 1186 | [100], [001] | p$_{55}$=p$_{44}$ | 0.1 | n$_o$ | [010] | [001], [100] | 32.971 |
| | | p$_{55}$=p$_{44}$ | 0.1 | n$_e$ | | [100], [001] | 19.555 |
| | | p$_{65}$=p$_{14}$ | 0.06 | n$_o$ | [001] | [010], [100] | not |
| | | p$_{65}$=p$_{14}$ | 0.06 | n$_o$ | | [100], [010] | not |
| 1230 | [010], [001] | p$_{24}$=-p$_{14}$ | 0.06 | n$_e$ | [100] | [010] | not |
| | | p$_{44}$ | 0.1 | n$_o$ | | [001], [010] | 29.557 |
| | | p$_{44}$ | 0.1 | n$_e$ | | [010], [001] | 17.530 |
| | | p$_{14}$ | 0.06 | n$_o$ | [001] | [100] | not |
| | | p$_{24}$=-p$_{14}$ | 0.06 | n$_o$ | | [010] | not |
| 2942 | [010], [100] | p$_{66}$ | 0.07 | n$_o$ | [001] | [010], [100] | not |
| | | p$_{66}$ | 0.07 | n$_o$ | | [100], [010] | not |

The change in the refractive index $n_3$ is given by

$$\Delta n_3 = \frac{1}{2}n_3^3 \left\{ p_{31}e_2 + p_{13}e_3 - \frac{(p_{44}e_4 - p_{41}e_2)^2}{(\frac{1}{n_1^2} - \frac{1}{n_3^2} + (p_{11} - p_{31})e_2 + (p_{13} - p_{33})e_3 - p_{14}e_4)} \right\}, \quad (3)$$

as well as by the relations $\frac{1}{n_1^2} - \frac{1}{n_3^2} \gg (p_{11} - p_{31})e_2 + (p_{13} - p_{33})e_3 - p_{14}e_4$ and

$$p_{31}e_2 + p_{13}e_3 \gg \frac{(p_{44}e_4 - p_{41}e_2)^2}{2\left[\frac{1}{n_1^2} - \frac{1}{n_3^2} + (p_{11} - p_{31})e_2 + (p_{13} - p_{33})e_3 - p_{14}e_4\right]}.$$

Eq. (3) may be simplified to the form

$$\Delta n_3 \approx 0.5 \times n_3^3 \{p_{31}e_2 + p_{13}e_3\}. \quad (4)$$

After considering the orientation of the displacement vector of acoustic wave, Eq.(4) becomes

$$\Delta n_3 \approx 0.5 \times n_3^3 \{-0.899 p_{31} + 0.457 p_{13}\}e. \quad (5)$$

Taking the values $p_{31}$=-0.15, $p_{13}$=0.14 and the relation $p_{ef} = \{-0.899 p_{31} + 0.457 p_{13}\}$ into account, one can arrive at $p_{eff}$=0.19 and $M_2$=150.32×10⁻¹⁵s³/kg. It can be seen that in case of the ultrasonic velocity achieving its lowest value (933.5m/s; to be compared with the value 1186m/s), the AOFM would increase drastically up to 150.32×10⁻¹⁵s³/kg (cf. with the previous value 32.971×10⁻¹⁵s³/kg). If we change the direction of the incident optical beam from the angle α=18° to α=180° with respect to the z axis (see Figure 3a) and change additionally the frequency of the acoustic wave from $f_a$=135×10⁶Hz (the collinear diffraction) to $f_a$=29×10⁹Hz, the AOFM would also change from $M_2$=240.7×10⁻¹⁵s³/kg through the value $M_2$=150.32×10⁻¹⁵s³/kg (for $k_i$ parallel to the y axis) up to $M_2$=243.4×10⁻¹⁵s³/kg. This effect is only owing to anisotropy of the refractive index $n_e$.

Let us now analyze the case of LTB crystals. When the incident optical wave is propagated along the [010] direction with $E_3$ polarization, while the acoustic wave (according to the Bragg condition, the acoustic frequency should be equal to $f_a$=2.8×10⁹Hz) is propagated in the (011) plane at the angle of 39° with respect to the z axis (the relevant projections of the unit displacement vectors being $X_z$=0.794, $X_y$=-0.606 and $X_x$=0), the optical indicatrix equation may be written as

$$(B_1 + p_{12}e_2 + p_{13}e_3)x^2 +$$
$$+(B_1 + p_{11}e_2 + p_{13}e_3)y^2 + \quad (6)$$
$$(B_3 + p_{31}e_2 + p_{33}e_3)z^2 + 2p_{44}e_4 yz = 1$$

Rewriting Eq. (6) in the proper coordinate system of the crystal, we get

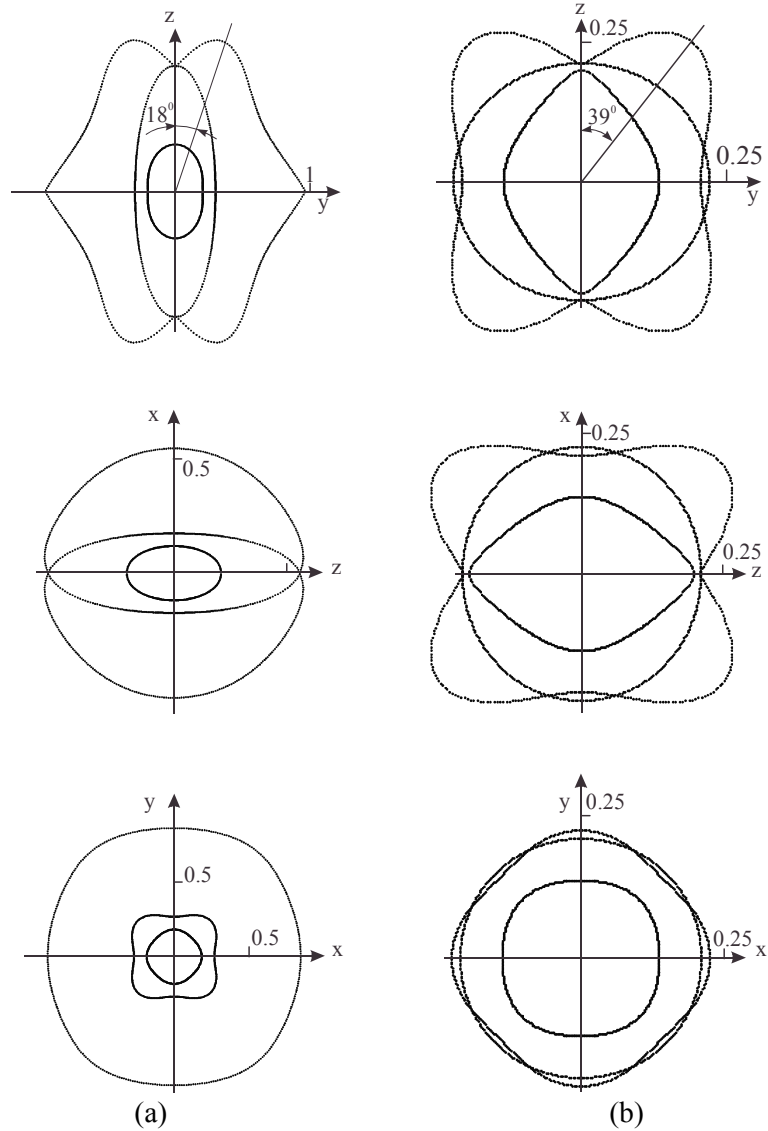

**Fig. 2.** Indicative surfaces of reciprocal acoustic wave velocities in the ABO (a) and LTB (b) crystals.

$$\left[B_1 + p_{12}e_2 + p_{13}e_3\right]X^2 + \left[B_1 + p_{11}e_2 + p_{13}e_3 + \frac{(p_{44}e_4)^2}{(B_1 - B_3 + (p_{11} - p_{31})e_2 + (p_{13} - p_{33})e_3)}\right]Y^2 + \\ \left[B_3 + p_{31}e_2 + p_{13}e_3 - \frac{(p_{44}e_4)^2}{(B_1 - B_3 + (p_{11} - p_{31})e_2 + (p_{13} - p_{33})e_3)}\right]Z^2 = 1 \quad (7)$$

Then the change in the refractive index $n_3$ reduces to

$$\Delta n_3 \approx \frac{1}{2}n_3^3\{p_{31}e_2 + p_{13}e_3\}. \quad (8)$$

With accounting for the orientation of the displacement vector of acoustic wave, Eq. (8) yields in

$$\Delta n_3 \approx \frac{1}{2}n_3^3\{-0.606\,p_{31} + 0.794 p_{13}\}e. \quad (9)$$

Since $p_{ef} = \{-0.606 p_{31} + 0.794 p_{13}\}$ and $p_{31}$=-0.24, $p_{13}$=-0.06, we obtain $p_{eff}$=0.1 and $M_2$=2.07×10$^{-15}$s$^3$/kg. If the ultrasonic velocity achieve its lowest value (3173m/s, not 4610m/s), the AOFM would increase by more than order of magnitude (2.07×10$^{-15}$s$^3$/kg; to be compared with the previous value 0.149×10$^{-15}$s$^3$/kg). Again, in case of changing direction of the incident optical beam from the angle $\alpha$=39º to $\alpha$=180º with respect to the $z$ axis (see Figure 3b), as well as simultaneously changing the frequency of the acoustic wave from $f_a$=732×10$^6$Hz (the collinear diffraction) to $f_a$=77.7×10$^9$Hz, the AOFM would evolve from $M_2$=2.35×10$^{-15}$s$^3$/kg through $M_2$=2.07×10$^{-15}$s$^3$/kg (for $k_i$ parallel to the $y$ axis) up to $M_2$=2.57×10$^{-15}$s$^3$/kg. Quite similar to the ABO crystals, this is owing to anisotropy of $n_e$.

The same value of the AOFM may be obtained when considering the second case of AO interaction in the LTB crystals. However, it is quite possible that the orientation of the acoustic wave vector intermediate between (011) and (101) planes could provide a less value of the ultrasonic wave velocity, when compare with the mentioned planes.

## Conclusion

In conclusion, one can notice that the ABO and LTB borate crystals manifest a high AOFM. The value $M_2$=243.4×10$^{-15}$s$^3$/kg for the ABO crystals is comparable, in the order of magnitude, with those typical for good AO materials such as TeO$_2$, for example. It is evident from the presented results that the most important criterion for the choice of crystals with a high value of AO parameter $M_2$ is the velocity of the acoustic wave. Moreover, we have shown that the propagation direction of the slowest acoustic wave does not necessarily coincide with the principal axes of the optical indicatrix ellipsoid. While changing the propagation direction of the acoustic wave (e.g., in the *xy*-plane), we have to take changing orientation of the displacement vector into consideration, the latter leading to the corresponding changes in the $p_{eff}$ parameter.

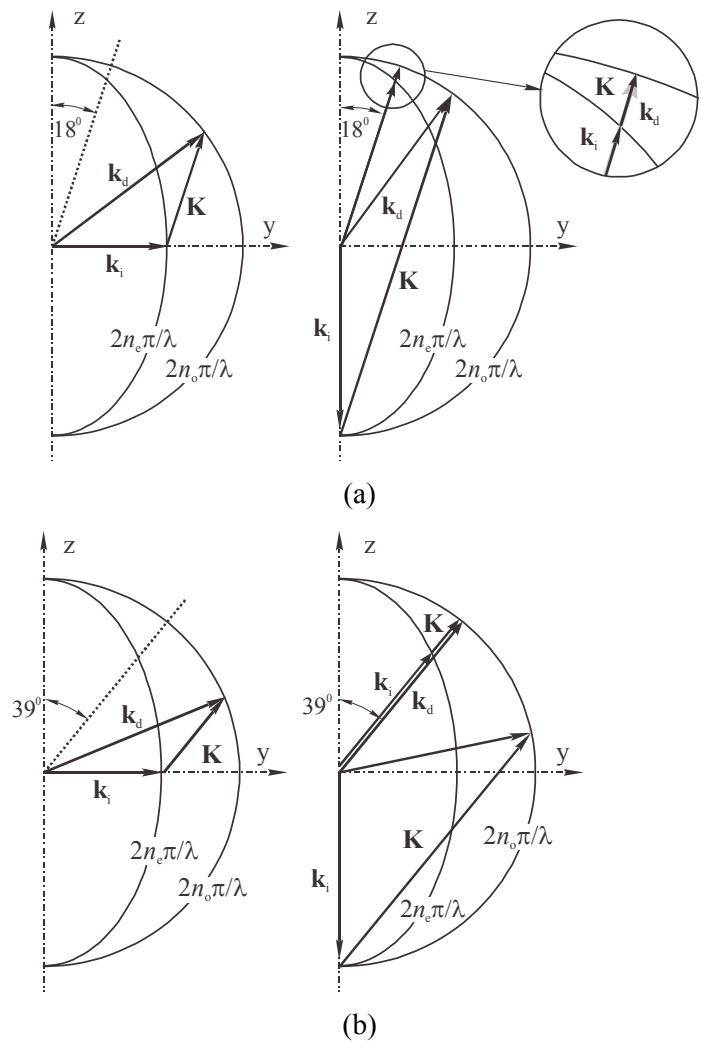

**Fig. 3.** Diagram of AO interaction with the slowest acoustic waves in the ABO (a) and LTB (b) crystals.

**Acknowledgement**

We would like to acknowledge the financial support of this study from the Scientific and Technology Centre of Ukraine under the Project N1712.